\documentclass[10.5pt,twocolumn,showpacs,showkeys,preprintnumbers,amsmath,amssymb,prb,superscriptaddress]{revtex4}
\usepackage[utf8]{inputenc}
\usepackage{amsmath}
\usepackage{amsfonts}
\usepackage{epsfig}
\usepackage{bm}
\usepackage{setspace}
\usepackage{booktabs} 
\usepackage{graphicx}
\usepackage[colorlinks=true, citecolor=red, linkcolor=blue, urlcolor=blue]{hyperref}
\begin{document}
\title{Relative abundances and enantiomerization energy of the chiral Cu$_{13}$ cluster at finite temperature}
\author{C\'esar Castillo-Quevedo}
\affiliation{Departamento de Fundamentos del Conocimiento, Centro Universitario del Norte, Universidad de Guadalajara, Carretera Federal No. 23, Km. 191, C.P. 46200, Colotl\'an, Jalisco, M\'exico}
\author{Carlos Emiliano Buelna-Garc\'ia}
\affiliation{Departamento de Investigaci\'on en Pol\'imeros y Materiales, Edificio 3G. Universidad de Sonora. Hermosillo, Sonora, M\'exico}
\affiliation{Organizaci\'on Cient\'ifica y Tecnol\'ogica del Desierto, Hermosillo 83150, Sonora, M\'exico}
\author{Edgar Paredes-Sotelo}
\affiliation{Departamento de Investigaci\'on en Pol\'imeros y Materiales, Edificio 3G. Universidad de Sonora. Hermosillo, Sonora, M\'exico}
\author{Eduardo Robles-Chaparro}
\affiliation{Departamento de Ciencias Qu\'imico Biologicas, Edificio 5A. Universidad de Sonora. Hermosillo, Sonora, M\'exico}
\author{Gerardo Mart\'inez-Guajardo}
\affiliation{Unidad Acad\'emica de Ciencias Qu\'imicas, \'Area de Ciencias de la Salud, Universidad Aut\'onoma de Zacatecas, Km. 6 carretera Zacatecas-Guadalajara s/n, Ejido La Escondida C. P. 98160, Zacatecas, Zac.}
\author{Jes\'us Manuel Quiroz-Castillo}
\affiliation{Departamento de Investigaci\'on en Pol\'imeros y Materiales, Edificio 3G. Universidad de Sonora. Hermosillo, Sonora, M\'exico}
\author{Aned de-Leon-Flores}
\affiliation{Departamento de Ciencias Qu\'imico Biologicas, Edificio 5A. Universidad de Sonora. Hermosillo, Sonora, M\'exico}
\author{Tulio Gaxiola}
\affiliation{Facultad de Ciencias F\'isico-Matem\'aticas de la Universidad Aut\'onoma de Sinaloa, 80010,  Sinaloa, M\'exico}
\author{Santos Jesus Castillo}
\affiliation{Departamento de Investigaci\'on en F\'isica, Universidad de Sonora, Blvd. Luis Encinas y Rosales S/N, 83000 Hermosillo, Sonora, M\'exico}
\author{Alejandro V\'asquez-Espinal}
\affiliation{Computational and Theoretical Chemistry Group Departamento de Ciencias Qu\'imicas, Facultad de Ciencias Exactas, Universidad Andres Bello, Republica 498, Santiago, Chile}
\author{Jose Luis Cabellos}\email[email:]{sollebac@gmail.com, josecabellos@uson.mx}
\affiliation{Departamento de Investigaci\'on en F\'isica, Universidad de Sonora, Blvd. Luis Encinas y Rosales S/N, 83000 Hermosillo, Sonora, M\'exico}
\date{\today}
\begin{abstract}
 This study reports the lowest energy structure of bare \href{https://youtu.be/pMcPvqIMgsw}{Cu$_{13}$ nanoclusters} as a pair of enantiomers  for temperatures ranging from 20 to 1200 K. Moreover, we compute the enantiomerization energy for the interconversion from $\mathcal{M}$  to  $\mathcal{P}$ structures in the chiral putative global minimum for temperatures ranging from 20 to 1300 K. Additionally, employing statistical thermodynamics and nanothermodynamics, we compute  the Boltzmann Probability for each particular isomer as a function of temperature. To achieve that, we explore the free energy surface of the Cu$_{13}$ cluster, employing a genetic algorithm coupled with density functional theory and statistical thermodynamics. Moreover, we discuss the energetic ordering of structures computed at the DFT level and compared to high level DLPNO-CCSD(T1) reference energies, and we present chemical bonding analysis using the {AdNDP} method in the chiral putative global minimum.  Based on the computed relative abundances, our results show that the chiral putative global minimum strongly dominates for temperatures ranging from 20 to 1100 K.
\end{abstract}
\maketitle
\section{Introduction}
Transition-metal (TM) nanoclusters have been widely studied due to their potential applications in catalysis~\cite{Chaves,https://doi.org/10.1002/anie.200901185,HENRY1998231}, photoluminescence~\cite{doi:10.1021/nl1026716}, photonics~\cite{Barnes2003}, magnetism~\cite{doi:10.1021/nl1026716}, chirality~\cite{doi:10.1021/jp101764q}, and the design of new materials~\cite{doi:10.1063/1.3054185,Majid2020}. Cu is a 3d TM with several oxidation states~\cite{Zhang2006,doi:10.1021/acs.chemrev.5b00482}, which explains its reactivity and confers many interesting physical and chemical properties~\cite{Zhang2006,doi:10.1021/acs.chemrev.5b00482}. Moreover, the high boiling point of Cu makes it compatible with high-temperature chemical reactions. Clusters are aggregates of atoms at the nanoscale size, which exhibit unusual physicochemical properties~\cite{C2CS15325D}. Cu clusters are particularly fascinating due to their applications in catalysis~\cite{doi:10.1021/acs.chemrev.5b00482}, light-emitting devices~\cite{C7NH00013H}, and nanotechnology~\cite{LIN2021100026}, despite presenting some challenges such as their easy oxidation~\cite{LIN2021100026}. The most stable structure of small Cu clusters has been investigated by density functional theory (DFT) studies~\cite{doi:10.1063/1.3187934,doi:10.1063/1.1436465,doi:10.1063/1.2431643,doi:10.1063/1.477313}  and considering the \emph{Jahn-Teller effect}~\cite{doi:10.1063/1.472939}. In the early 2000s, Poater et al.~\cite{doi:10.1021/jp054690a} characterized neutral copper clusters (Cu$_n$ n = 1-9) using computed chemical reactivity descriptors within the DFT framework. Later, atomic structures and reactivity descriptors of Cu$_n$CO (n = 1,9) were computed and discussed~\cite{doi:10.1021/jp054690a}. Calaminici et al. reported the structure of neutral and anionic Cu$_9$ clusters, employing DFT~\cite{doi:10.1021/ct600358a}. Moreover, in previous combined theoretical-experimental studies, the computed removal energies were compared with the measured photoelectron spectra in anionic Cu$_n$ (n= 9,20) clusters~\cite{doi:10.1063/1.3300128}, and later, the optical absorption of small Cu clusters was presented~\cite{doi:10.1063/1.3552077}. Based on their geometry and electronic structure, atomic clusters could be characterized by magic numbers~\cite{BINNS20011,Chaves,RevModPhys.65.677,MARTIN1996199}  that form highly symmetric structures; for example, icosahedron (ICO) and cuboctahedron (CUB) shapes~\cite{Chaves}. From the geometrical point of view, the first magic number that appears is 13. Experimental studies have found magic TM$_{13}$ clusters for Fe and Ti, amongst others~\cite{doi:10.1063/1.479268,Chaves}. Previous theoretical studies based on the empirical potential of Cu$_{13}$ clusters showed that the low-energy structures were the icosahedron and the cuboctahedron~\cite{doi:10.1063/1.1429658}; those structures consist of a central atom surrounded by 12 Cu atoms. In contrast, Guvelioglu et al.~\cite{PhysRevLett.94.026103}, within the framework of DFT, found that the lowest energy structure of Cu$_{13}$ is the double-layered structure, and in the same year Itoh et al.~\cite{doi:10.1142/S0217979205031080}  reported a similar double-layer structure as the putative global minimum. Yang et al. explored the structural evolution of Cu$_n$ (n = 8-20) anions and found platelike structures~\cite{doi:10.1063/1.2150439}. Later, larger Cu$_n$ (n = 20-30) clusters were investigated, and it was found that the structures are based on a 13-atom icosahedral core~\cite{doi:10.1063/1.3689442}. In the previous studies, Cu$_{13}$ was investigated because it was found to have an icosahedral structure that has a high percentage of edge and corner sites and high-index facets, resulting in increased catalytic activities~\cite{Riguang,Shuangjing,Takahashi}. In most cases, low-energy Cu clusters have preferentially lower symmetry structures~\cite{Chou_2013}, although some present distorted structures~\cite{Chou_2013,PhysRevB.58.3601}. Although there are many studies on Cu clusters, the chirality of Cu$_{13}$ clusters has not been discussed. In general, chirality plays a decisive role in biological activity and life processes.~\cite{Ebeling2018}. Remarkably, chiral nanoclusters have attracted attention because they have applications in chiral materials with novel properties~\cite{https://doi.org/10.1002/anie.201402488,C7CC08191J,molecules26133953}. Recently, theoretical studies on beryllium boron cluster Be$_4$B$_8$ at the DLPNO-CCSD(T1) theoretical level found chiral structures as the lowest energy structures~\cite{molecules26133953}. Previous theoretical studies on PtPd co-doped silicon clusters reported chiral and fluxional low-energy structures~\cite{LV2017873}. Recently, Kong et al. reported propeller-like chiral AIE copper (I) clusters with exciting properties~\cite{https://doi.org/10.1002/anie.201915844}. However, clusters’ properties depend on their putative global minimum and low-energy structures, considering achiral and chiral structures. Hence, we need to know the distributions of isomers at different temperatures~\cite{Buelna,molecules26133953,Baletto}. The lowest- and low-energy geometries, composition, and temperature of the ensemble determine all the properties of a cluster at temperature T~\cite{Buelna,RAO201350,Baletto}, i.e., its electronic, structural, thermodynamics, vibrational, and optical properties, as well as its chemical reactivity. Moreover, the atomic structure is the first level at which it is possible to manipulate the macroscopic properties of a cluster~\cite{Chaves}.

In this study, intending to elucidate the lowest- and the low-energy structures of neutral Cu$_{13}$ clusters at finite temperature, we explored their free energy surface by employing a genetic algorithm coupled to DFT, statistical thermodynamics and nanothermodynamics. We computed the probability of occurrence of each particular isomer, employing Boltzmann probabilities for temperatures ranging from 20 to 1500 K. Our findings show that the putative global minimum is a chiral structure  at room temperature. Moreover, we computed the transition state (TS), i.e., the enantiomerization energy for temperatures ranging from 20 to 1300 K, for interconversion of a pair of enantiomers (Plus, $\mathcal{P}$, and Minus, $\mathcal{M}$). Our computations showed that enantiomerization barriers led to persistently chiral structures and enabled the complete separation of enantiomers at room temperature~\cite{Yang,molecules26133953}. The remainder of the manuscript is organized as follows. Section 2 provides the computational details and a brief overview of the theory and the algorithms used. The results and discussion are presented in Section 3. We discuss the low-energy structures, the effect of the DFT functionals on the energetic ordering of isomers and comparison to DLPNO-CCSD(T1)~\cite{doi:10.1021/acs.jpca.9b05734} reference energies; T$1$ diagnostic is presented. The effect of the symmetry number on the Gibbs free energy and on the thermal populations at temperature T; and the origin of the slight 0.41 kcal/mol energy differences are investigated. We analyze the interconversion energy barrier between $\mathcal{P}$ and $\mathcal{M}$ enantiomers, the effects of the temperature in the energy barriers, and the thermal population. Conclusions are given in Section 4.
\section{Theoretical Methods} 
\subsection{Global Minimum Search and Computational Details}
All geometrical structures are optimized locally without imposing  any symmetry, the self-consistent-field procedure was performed with a
convergence criterion of 10$^{-6}$ a.u. The energy,  maximum force  and maximum displacement convergence were set to
10$^{-6}$ Ha, 0.002 Ha/\AA~ and 0.0005~\AA~respectiveley. All calculations
were performed using the Gaussian suite code~\cite{gaussian} employing the
Becke's hybrid three-parameter~\cite{Becke2,Becke}
exchange-correlation functional in combination with the Perdew and Wang GGA functional PW91,~\cite{pw912,pw91}
this combination is known as B3PW91 exchange-correlation functional. 
The B3PW91 has been employed in other studies of reactivity in copper clusters with a good performance~\cite{avelino,pablo},
It is worth point out that hybrid functionals including a portion of Hartree-Fock exchange have shown
a superior performance~\cite{avelino,cramer,Cohen}.
For the basis set, we employed the 
The LANL2DZ basis set is used for transitions metals due to its low computational cost~\cite{Hay}   
With the aim to refine the optimization and the energies, we used Ahlrichs-type triple-$\zeta$ quality
extended valence def2TZVP basis~\cite{Ahlrichs,Zeng2010} that is more accurate for transition metals~\cite{Tayyibe}
despite it has a considerably higher computational cost.~\cite{Matczak} In this study the dispersion corrections are taking
into account trough the D3 version of Grimme’s dispersion~\cite{grimme} as it is implemented in Gaussian code.
In a previous work, the effect of the dispersion corrections on the structural and energetic properties of
the pure BeB clusters were studied founding that the energetic ordering of isomers can change when
the dispersion is taking into account~\cite{Buelna,molecules26133953}. Transition states are discarded
through a vibrational analysis, making sure that the minimum energy structure is a true structure minimum energy.
Calculation of Gibbs free energy properties of Cu$_{13}$ cluster
requieres an exhaustive and systematic sampling of the free energy surface
with the aim to find all possible low-energy structures.~\cite{Buelna,Heiles} First of all, 
the search of the global minimum in atomic clusters is a complicated task due mainly that degrees of
freedom of a molecule increases with the number of atoms; as a consequence, the number of local minimum
increase exponentially with the number of atoms, moreover the calculated total energy of the cluster
requisites high level of quantum mechanical methodology to produce a real energy. In spite of that,
several algorithms coupled with DFT to search the lowest energy structures on the potential
energy surface of atomic clusters has been employed until nowadays,
as kick methodology~\cite{vargas,ravell,pan,cui,alba,cui2,Vargas-Caamal2015,florez,grande,rosa},
genetic algorithms~\cite{Buelna,Dong,Guo,Dong2,Alexandrova} Our computational procedure to
elucidate the lowest energy consisted of employing a
hybrid genetic algorithm called~\emph{GALGOSON}~\cite{Buelna,molecules26133953}
\emph{GALGOSON}
employ a multi-step search strategy where in the first step it makes optimizations employing LANL2DZ basis set and
in second step-refine it employs the def2TZVP basis set, the creation of the initial population take into account the
2D and 3D structures~\cite{Dong,grande} with an initial population of 650 random structures for the Cu$_{13}$ cluster
and the criterion to stop the algorithm is until the five generations converged.
This methodology based on previous works~\cite{kcabellos,Hernandez,Guo,grande} consists of a multi-step
approach (cascade) to efficiently sample the potential/free energy surface coupled to the \emph{Gaussian} code~\cite{gaussian}.
Chemical bonding was examined using the Adaptive Natural Density Partitioning (AdNDP) method~\cite{B804083D,molecules26133953}. AdNDP analyses the first-order reduced density matrix and recovers Lewis bonding (1c–2e or 2c–2 e, i.e., lone pairs (LPs) or two-center two-electron bonds) and delocalized bonding elements (associated with the concept of electron delocalization).

In this study, the fundamental thermodynamic properties such as enthalpy H, and entropy S and Gibbs
free energy dependent on temperature were computed within the framework of
nanothermodynamics~\cite{Buelna,doi:10.1021/ja073129i,molecules26133953,Dzib,C4SC00052H,https://doi.org/10.1002/chem.201200497}
through the partition function Q  given in Equation~\ref{partition1}, and described in refs.~\cite{Buelna,D1SC00621E,doi:10.1021/ja073129i} or any standard text relating
to thermodynamics~\cite{mcquarrie1975statistical,hill1986introduction}
\begin{equation}
\displaystyle
Q(T)=\sum_{i}g_i~e^{{-\Delta{E_i}}/{K_BT}}
\label{partition1}
\end{equation}
In Eq.~\ref{partition1}, $g_i$ is the degeneracy factor, $k_{\textup{B}}$ is the Boltzmann constant, $T$ is the temperature, and
${-\Delta{E_i}}$ is the total energy of a cluster.~\cite{Buelna,Dzib,mcquarrie1975statistical}.
The total partition function Q is the product of the qtrasn, qrot, qvib, and qelec~\cite{Buelna,D1SC00621E,molecules26133953} computed
under the rigid rotor, harmonic oscillator, Born-Oppenheimer, ideal gas, and particle-in-a-box approximations~\cite{Buelna,molecules26133953}
The thermal populations P(T) at absolute temperature T or the so called  probability of a particular isomer is computed with Equation~\ref{boltzman}:
\begin{equation}
\centering 
\displaystyle
P(T)=\frac{e^{-\beta \Delta G^{k}}}{\sum e^{-\beta \Delta G^{k}}}\label{boltzman}, 
\end{equation}
where $\beta=1/k_{\textup{B}}T$, and $k_{\textup{B}}$ is the Boltzmann constant, $T$ is the absolute temperature, $\Delta G^{k}$ is the Gibbs free energy
of the $k^{th}$ isomer. Equation~\ref{boltzman} establishes that the distribution of molecules
among energy levels is a function of the energy and temperature~\cite{MENDOZAWILSON2020112912,Buelna,molecules26133953}.

\section{Results and Discussion}
\subsection{The lowest energy structures and energetics}
\begin{figure*}[ht!]
\begin{center}  
  \includegraphics[width=\textwidth]{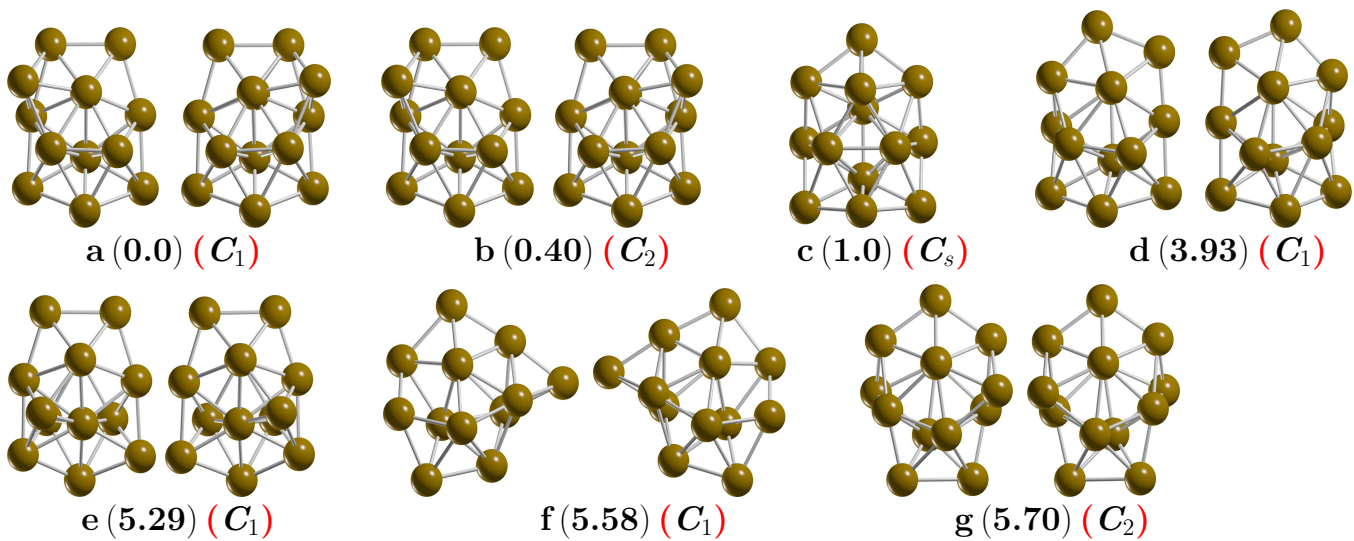}
  \caption{(Color online) The chiral lowest energy structure is depicted in a),  the chiral low-energy structures are depicted in b,d,e,f,g whereas the achiral structure is depicted in c).  Those structures were  optimized at the  B3PW91-D3/def2TZVP level of theory The first letter indicates the isomer, the relative Gibbs
    free energies in kcal/mol appear in round parentheses computed at 298.15 K, and the point group symmetry in red round parentheses. The isomers, represented in a) and b), differ only in molecular symmetry. The isomer with C2 symmetry has a Gibbs free energy equal to less than the non-symmetric C$_1$ isomer. The RMSD between the isomer with symmetry C$_1$ and the isomer with symmetry C$_2$ is 0.0014. The atomic XYZ coordinates are given in Appendix~\ref{appendix:b}.}
\label{geometry_gibbs}
\end{center}    
\end{figure*}
The most important low-energy structures of a neutral Cu$_{13}$ cluster optimized at the B3PW91-GD3/def2TZVP
level of theory found in this study are shown in Figure~\ref{geometry_gibbs}. At room temperature, the isomers depicted in Figure~\ref{geometry_gibbs}
contributed to 94{\%} of the molecular properties in a Boltzmann ensemble; thus, almost all molecular properties were due to those isomers.
Additionally, they are chiral structures. The putative chiral global minimum is depicted in Figure~\ref{geometry_gibbs}a  with symmetry C$_1$.
These are bilayered structures composed of a shared pentagonal bipyramid interspersed with a distorted hexagonal ring with a Cu atom capping one of its
faces and two Cu atoms capping the other face of the hexagonal ring, in good
agreement with similar structures~\cite{peter,PhysRevLett.94.026103,Limbu_2019}.
The pentagonal bipyramid interspersed with the hexagonal ring is built with 12 Cu atoms. One more Cu atom caps
the pentagonal bipyramid; this capping Cu atom is responsible for the chirality of the Cu$_{13}$ cluster.
Our calculated Cu-Cu bond length on the putative chiral global minimum is 2.432~\AA,~in good agreement with other
reported DFT calculations of a Cu-Cu dimer of 2.248~\AA~\cite{Galip} and also with an experimental bond length of 2.22~\AA~\cite{kabir,Linus},
just slightly above 5.3{\%} of the experimentally determined value. The calculated vibrational frequency of Cu$_{13}$ was 60 cm$^{-1}$,
whereas the computed vibrational frequency of the Cu-Cu dimer was 245 cm$^{-1}$, again in good agreement with
the experimental value of 265 cm{$^{-1}$}~~\cite{Linus}. We also explored the higher multiplicity of quartets and found that the lowest energy
structure lay 20.5 kcal/mol above the doublet putative chiral global minimum energy structure.
The second structure that was higher in free energy lay at 0.41 kcal/mol at room temperature;
it was also a bilayered structure, similar to the putative global minimum, but with C$_2$ symmetry.
Iwasa et al.~\cite{doi:10.1021/acs.jpca.8b08868} reported a similar double-layer structure as a putative
global minimum with C$_2$ symmetry, but without taking into account the temperature. One of our previous studies
showed that these tiny Gibbs free energy differences are derived from rotational entropy~\cite{molecules26133953}. The C$_1$ and C$_2$
symmetry clusters adopted a hollow layered structure. The following higher energy isomer lay at 1.0 kcal/mol at room temperature and was
an achiral buckled-biplanar (BBP) structure with C$_s$ symmetry, which agrees with previous work~\cite{gallego}. At room temperature,
its contributions to the molecular properties were less than 6{\%}. The average bond length on isomer BBP was 2.432~\AA,~similar to
the average bond length of the chiral putative global minimum. Next, higher energy structures lay 3.9 kcal/mol above the chiral putative global minimum,
and their average bond length was 2.444~\AA,~ slightly larger than the average bond length of 2.432~\AA~of the putative global minimum.
This also appeared as a bilayered chiral structure with a shared hexagonal bipyramid interspersed with a hexagonal bipyramid. The following higher
energy structure lay 5.29 kcal/mol above the putative minimum global. It was a bilayered structure consisting of 12 atoms, with 1 atom capping one of its faces.
It is depicted in Figure~\ref{geometry_gibbs}e. Structures located at higher energy than 5.5 kcal/mol above the putative global minimum are depicted
in Figure~\ref{geometry_gibbs}f,g. Those structures also adopted a layered structure with no interior atoms, with similar morphology to that of
low-energy isomers. These two structures did not contribute to the molecular properties in the studied temperature range. The Cu$_{13}$
cluster low-energy structures preferentially adopted morphologies of bilayered structures rather than highly symmetric 3D structures.
In contrast, Au$_{13}$ clusters prefer planar structures due to relativistic effects~\cite{Chou_2013}; therefore, further studies are
needed to investigate why bilayered structures in the Cu$_{13}$ cluster are energetically preferred. For the Cu$_{13}$ cluster,
the icosahedron structure is not energetically favorable in the temperature range examined, which is consistent with previous work
where the authors did not take into account the temperature~\cite{Chaves}. In this study, the icosahedron structure was
located at 24.6 kcal/mol above the putative global minimum at room temperature.
\begin{figure}[ht!]
\begin{center}  
\includegraphics[scale=0.30]{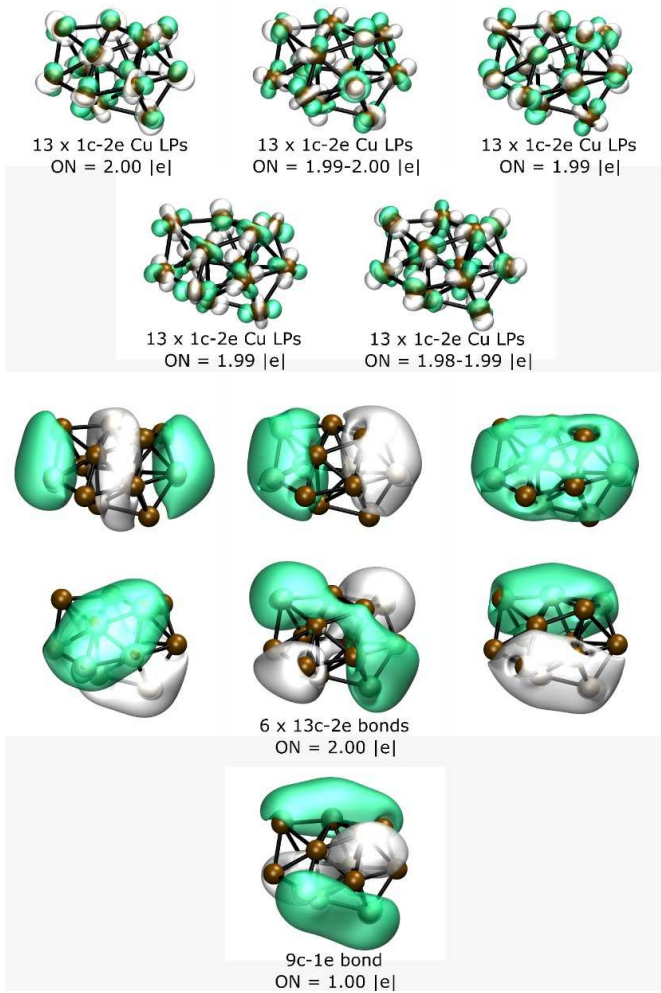}
\caption{(Color online) Results of the AdNDP analysis of the lowest-energy chiral isomer of the Cu$_{13}$ system.} 
\label{adnp}
\end{center}
\end{figure}
To get an idea of the bonding situation in the chiral
putative global minimum structure, we performed an AdNDP analysis; the results are shown in Figure~\ref{adnp}. This analysis revealed the presence of
5 sets of 13 1c-2e bonds with occupation numbers (ONs) between 1.98 and 1.99 |e|, i.e., lone pairs (LPs) corresponding to the fully
filled 3d shell in each Cu atom. The bonding in this cluster was then due to the 4s shell electrons for which the bonding pattern, as revealed by
AdNDP, consisted of 6 sets of 13c-2e completely delocalized bonds, plus a 9c-1e bond corresponding to the unpaired electron which, as shown in Figure~\ref{adnp},
was mostly delocalized in the peripheral atoms of the cluster.

\subsection{Energetics}
Temperature drastically affects the Gibbs free energy of the isomers; therefore, in a molecular ensemble (collection),
the energetic ordering of isomers changes. Besides, from a theoretical point of view, the energetic ordering can also change
when computing energies using different levels of theory~\cite{Buelna,PhysRevB.56.7607}. To wit, energy computed at different methods yield different
energies due mainly to the functional and basis-set employed,~\cite{Buelna}, so the energetic ordering change;
consequently, the probability of occurrence and the molecular properties will change. Moreover, we compute the
$\mathcal{T}_1$ diagnostic to determine if the computed DFT energies are properly described by a single reference method  or contain a
multireference character.

To gain further insight into
the energetic ordering of the low-lying isomers, we optimized the low-lying energy structures employing three more DFT functionals:
TPSS~\cite{PhysRevLett.91.146401}, PBE~\cite{perdew1996}, and BP86~\cite{Becke} with def2-TZVP basis set, and single point (SP) relative energies
computed employing the domain-based local pair natural orbital coupled-cluster theory (DLPNO-CCSD(T)), with TightPNO setting~\cite{doi:10.1021/acs.jpca.9b05734},
and with and without ZPE energy correction. The purpose was to ascertain the origin of the slight 0.41 kcal/mol differences
(below the chemical accuracy of 1 kcal/mol) in the relative Gibbs free energy (Table~\ref{energia1})
and that these are not due to numerical  errors, algorithmic approximations, integration grids, or functional and basis
set dependence, to name a few.
In Table~\ref{energia1}, lines first through the fourth show the relative Gibbs free energies computed at
B3PW91-D3/def2-TZVP, TPSS-D3/def2-TZVP, PBE-D3/def2-TZVP, and BP86-D3/def2-TZVP, respectively.
DLPNO-CCSD(T)) relative energies, with and without ZPE correction, are shown in the fifth and sixth line of Table~\ref{energia1},
electronic with zero-point energy and electronic energy, at the B3PW91-D3/def2-TZVP level of theory are shown in the seventh and eighth lines
of Table~\ref{energia1}, and line ninth shows the point group symmetry, the $\mathcal{T}_1$  diagnostic for each isomer
is shown in line tenth  of Table~\ref{energia1},  our results confirm that the computed $\mathcal{T}_1$ diagnostic values are slightly above 
the recommended threshold of 0.02~\cite{https://doi.org/10.1002/qua.560360824,molecules26133953}; So, these values suggest
that multireferential studies have to be carried out for Cu$_{13}$ cluster.
At the CCSD(T) theoretical level, the lowest-energy structure is the pair of enantiomers with symmetry C$_2$;
despite that, at finite temperature, the lowest energy structure is the pair of enantiomers with symmetry C$_1$.
A more detailed analysis of the results in Table~\ref{energia1} shows that the relative electronic energy of the four chiral
low-energy isomers labeled in Table~\ref{energia1} (a, b, c, d), with symmetry C$_1$, C$_1$, C$_2$, and C$_2$, respectively,
is zero, considering the ZPE; also the relative electronic energy is zero.
In contrast, the relative Gibbs free energy at 298.15 K shown in the fifth column is 0.41 kcal/mol.
The relative Gibbs free energy at 298.15 K for the TPSS, PBE, BP86 DFT-functions,
between the putative global minimum and the second isomer, is also 0.41 kcal/mol. This Gibbs free energy difference does not
depend on the functional employed, as shown in Table~\ref{energia1}. At temperature T=0,
the total energy of an isomer is the electronic energy plus ZPE ($\varepsilon_0+ZPE$). If the temperature increases,
entropic effects start to play, and Gibbs’s free energy determines the global minimum. At any temperature T, the isomers,
represented in Figure~\ref{geometry_gibbs} and  Figure~\ref{geometry_gibbs}b, differ only in molecular symmetry. The isomer with C$_2$
symmetry has a Gibbs free energy equal to RTIn$(\sigma)$ less than the non-symmetric C$_1$ isomer. Here R is the universal gas constant,
T temperature, and $\sigma$ is the symmetry number. The symmetry number appears in the denominator of the
molecular rotational partition function~\cite{Fernandez-Ramos2007,Buelna,molecules26133953,doi:10.1021/jp110434s}.
This implies that the less symmetric isomers at finite temperature are more thermodynamically stable than the more symmetric ones due to the energy factor
given by RTIn$(\sigma)$. The factor becomes zero at T=0 and increases linearly with temperature.
\begin{figure}[ht!]
\begin{center}  
  \includegraphics[scale=0.80]{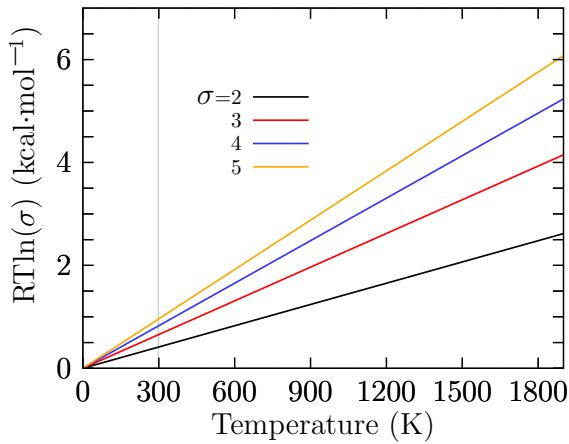}
\caption{(Color online) The difference of the Gibbs free energy computed with and without symmetries is given by a factor of RT$\ln(\sigma)$; in this factor,  R is the universal gas constant, T, the temperature, and $\sigma$ is the symmetry number (The factor is similar to the enantioselectivity. These values, at 298.15 K,  are in good agreement with the value 0.41 kcal/mol shown in the Table~\ref{energia1}.} 
\label{rot2}
\end{center}
\end{figure}
Figure~\ref{rot2} shows the  factor as a function of temperature and for different symmetry numbers. Moreover the $\mathcal{T}_1$ diagnostic
indicate that a multireference c
For our optimized low-energy isomers with C$_2$ symmetry, the symmetry number is 2, so the Gibbs free energy at 298.15 K with and without symmetry
will differ by 0.41 kcal/mol regardless of  the DFT method. This value is higher at high temperatures and with higher symmetry numbers.
For example, the benzene molecule with D$_{6h}$ symmetry has a symmetry number 12, the Gibbs free energy at 298.15 K with and without symmetry will differ
by 1.47 kcal/mol, which is greater than the chemical precision. Here, we call this the effect of the symmetry number on the Gibbs free energy and on
the thermal populations at finite temperature; the symmetry number appears when identical atoms are considered indistinguishable and are determined
solely by the point group symmetry of the molecule. We emphasize the importance of symmetry in calculating thermal populations at absolute
temperature T or the so called population probability or relative populations, hence the molecular properties. For example, the melting temperature
for a symmetrical molecule is higher than a non-symmetrical molecule; moreover, the activation energy barrier could be higher when we
consider a non-symmetrical molecule in the calculation of the transition state. The energy computed at different theoretical levels influences
the energy distribution of the isomers and, as a consequence, the Boltzmann weights. For the four DFT functionals used in this study,
the energy ordering is preserved, although differences in the energy between the isomers occur; for each DFT functional, the main contributors
to any molecular property of Cu$_{13}$ are always the chiral isomers depicted in Figure~\ref{geometry_gibbs}a.
\begin{table*}[!ht]\centering
  \caption{Relative energy in kcal/mol of the low-lying isomers depicted in Figure~\ref{geometry_gibbs} labeled from a to f,
    employing various density functionals:  B3PW91, TPSS, PBE, and BP86,  with the def2TZVP basis set.
    CCSD(T) employing  the domain-based local pair natural orbital coupled-cluster theory (DLPNO-CCSD(T)), with TightPNO setting,
   with zero point energy (DLPNO-CCSD(T){\normalsize{$+\mathcal{E}_{\mathrm{ZPE}}$})}. Electronic energy with $\mathrm{ZPE}$ {(\normalsize{$\mathcal{E}_0+\mathcal{E}_{\mathrm{ZPE}}$)}}, Electronic energy {(\normalsize{$\mathcal{E}_0$)}}, point group symmetry, and \normalsize{$\mathcal{T}_1$ Diagnostic.}}
\label{energia1}
 \begin{tabular}{@{\extracolsep{0.2pt}} ll l l lll lll l }
 \\[-1.8ex]\hline 
 \hline \\[-1.8ex] 
  &   &  \multicolumn{9}{c}{\normalsize{Isomers}}\\
\cline{2-11}\\ [-1.8ex] 
\multicolumn{1}{c} {{{ \normalsize{ Level }}}}  & \multicolumn{1}{l}{$i_a$} & \multicolumn{1}{c}{$i_a$} & \multicolumn{1}{c}{$i_b$ } & \multicolumn{1}{c}{$i_b$}& \multicolumn{1}{c}{$i_c$}& \multicolumn{1}{c}{$i_d$}  &\multicolumn{1}{c}{$i_d$} &\multicolumn{1}{c}{$i_e$}&\multicolumn{1}{c}{$i_e$} & {$i_{f}$}  \\
\hline \\[-1.8ex]
$\Delta G$(B3PW91)                                & 0.0  &  0.0  & 0.3953  & 0.4091  &  0.9946  &  3.9338   &  3.9357  & 5.2967 & 5.2967 & 5.5728 \\
$\Delta G$(TPSS)                                  & 0.0  &  0.0  & 0.4010  & 0.4016  &  0.7248  &  5.6017   &  5.6029  & 5.5477 & 5.5477 & 7.5438\\
$\Delta G$(PBE)                                   & 0.0  &  0.0  & 0.4145  & 0.4217  &  0.5491  &  3.8478   &  3.8491  & 5.5283 & 5.7284 & 5.7310 \\
$\Delta G$(BP86)                                  & 0.0  &  0.0  & 0.4079  & 0.4035  &  0.8634  &  3.5611   &  5.4291  & 5.4291 & 5.2817 & 7.2798 \\
DLPNO-CCSD(T)                                       & 0.0  &  -0.1298 &  -0.2154  &  -0.2144  &  0.6827  & 5.0246  &   5.8771 & & & \\
DLPNO-CCSD(T)\normalsize{$+\mathcal{E}_{\mathrm{ZPE}}$}& 0.0  &  -0.1298 &  -0.2216  &  -0.2170  &  0.7084  & 4.9022  &   5.7886 & & & \\
\normalsize{$\mathcal{E}_0+\mathcal{E}_{\mathrm{ZPE}}$}& 0.0  &  0.0      &  0.0     & 0.0       &  0.9262  & 4.5877  &   4.5877 & 5.6124 & 5.6124 & 6.3327\\
\normalsize{$\mathcal{E}_0$}                        & 0.0  &  0.0      &  0.0     & 0.0       &  0.9002  & 4.7096  &   4.7098 & 5.7013 & 6.4969 & 6.4966\\
{Point group symmetry}                              & {\normalsize{\emph{C$_{1}$}}}  &  {\normalsize{\emph{C$_{1}$}}} &  {\normalsize{\emph{C$_{2}$}}}   & {\normalsize{\emph{C$_{2}$}}}
& {\normalsize{\emph{C$_{s}$}}}         &  {\normalsize{\emph{C$_{1}$}}} & {\normalsize{\emph{C$_{1}$}}}   & {\normalsize{\emph{C$_{1}$}}} & {\normalsize{\emph{C$_{1}$}}}& {\normalsize{\emph{C$_{1}$}}}   \\
\normalsize{$\mathcal{T}_1$ Diagnostic}             & 0.023  &  0.023  &  0.023   & 0.023  & 0.023   & 0.023  & 0.023 & 0.023  & 0.023 & 0.023\\
\hline \\[-1.8ex]
\end{tabular}
\end{table*}
\subsection{Structures and Stability at Finite Temperature.}
\begin{figure}[ht!]
\begin{center}  
\includegraphics[scale=0.85]{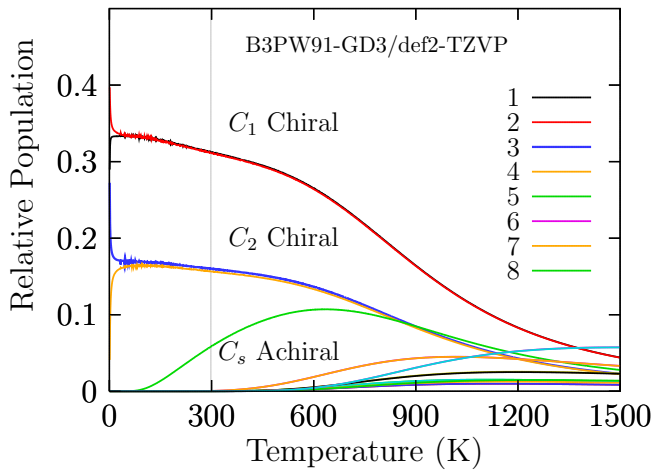}
\caption{(Color online) Probability of occurrence for all isomers at temperatures ranging from 20 to 1500 K. The red and black solid lines depict the probability occurrence of the putative chiral global minimum with symmetry C$_1$ and strongly dominate in all ranges of temperatures.
  The bulk melting temperature of copper is 1358 K~\cite{Grigoryan}; thus, our results below this temperature are consistent. Probability of occurrence  with  DFT  functionals: TPSS, PBE, and BP86 are displayed  in Figure~\ref{popu2}.} 
\label{popu}
\end{center}
\end{figure}
In chemistry, physics, and biology, the lowest energy structure and all the low-energy structures near the global minimum are crucial because all molecular properties are statistical averages derived from the ensemble of molecular conformations~\cite{molecules26133953}.
The probability of occurrence of each particular isomer is depicted in Figure~\ref{popu} for the Cu$_{13}$ cluster. It was determined by employing Equation~\ref{boltzman} and temperatures ranging from 20 to 1500 K at the B3PW91-D3/def2TZVP level of theory. Figure~\ref{popu} shows
the probability of occurrence considering all chiral and achiral structures. The analysis of these results led to
an interesting observation. The pair of enantiomers that appeared as the putative global minimum at temperature 0 K was strongly dominant in the
temperature range from 20 to 1500 K. Moreover, there were no solid-solid transformation points in any temperature range,
which means no interchange of dominant low-energy structures at high temperatures. A closer inspection of Figure~\ref{popu} shows
that the decay of probability of occurrence of the pair of enantiomers with symmetry C$_1$, depicted by a red  solid line,
is closer to linear rather than exponential for temperatures ranging from 20 to 600 K. Above 600 K and up to 1500 K, the decay is exponential.
At 300 K, the chiral structure has a probability of 32{\%}, whereas the second isomer located 0.4 kcal/mol above the putative global minimum has
a probability of 16{\%}. The above discussion shows that all molecular properties of the Cu$_{13}$ cluster are attributed to the chiral
putative global minimum at absolute temperature zero.
The probability of occurrence of the chiral putative global structures, is 
depicted by blue and yellow solid lines in Figure~\ref{popu}. The probability of occurrence for structures with C$_1$ and C$_2$
symmetries showed similar behaviors but different values; even so, the molecular properties are attributed to only one pair of enantiomers with symmetry C$_1$.
The probability of occurrence for the achiral structure, which is shown in Figure~\ref{geometry_gibbs}c,
located 1 kcal/mol above the putative global minimum, is depicted by
a green solid line in Figure~\ref{popu}; it started to increase around a temperature of 120 K, and at room temperature, it  is  has a probability of 5{\%}. At 700 K, the highest probability of occurrence was reached, corresponding to 12{\%}; above this temperature, up to 1500 K, it started to decrease. Note that above 800 K and up to 1200 K, the achiral structure with C$_s$ symmetry and the putative global minimum structures with C$_2$ symmetries coexisted.
Interestingly, the Boltzmann ensemble was composed of an equal mixture of  $\mathcal{M}$ and $\mathcal{P}$ enantiomers; thus,
the chiral properties were null in all ranges of temperature, i.e., the Cu$_{13}$ cluster did not exhibit properties such as
vibrational/electronic circular dichroism.  In ranging temperatures from 1200 to 1500 K, all isomers coexist with less than ten percent probability.
To wit, all isomers are equally populated for hot temperatures or when the temperature increases to large values.
The bulk melting temperature of copper, 1358 K~\cite{Grigoryan}; So, we must consider that the anharmonic effects become strong
at high temperatures~\cite{molecules26133953}. From the thermal population, we consider the entropic-temperature term has a small effect on the Cu$_{13}$
cluster distribution of isomers on the scale of temperature, as shown in Figure~\ref{popu}.

\subsection{Enantiomerization Energy Barrier at Finite Temperature.}
\begin{figure*}[ht!]
\begin{center}  
  \includegraphics[scale=0.75]{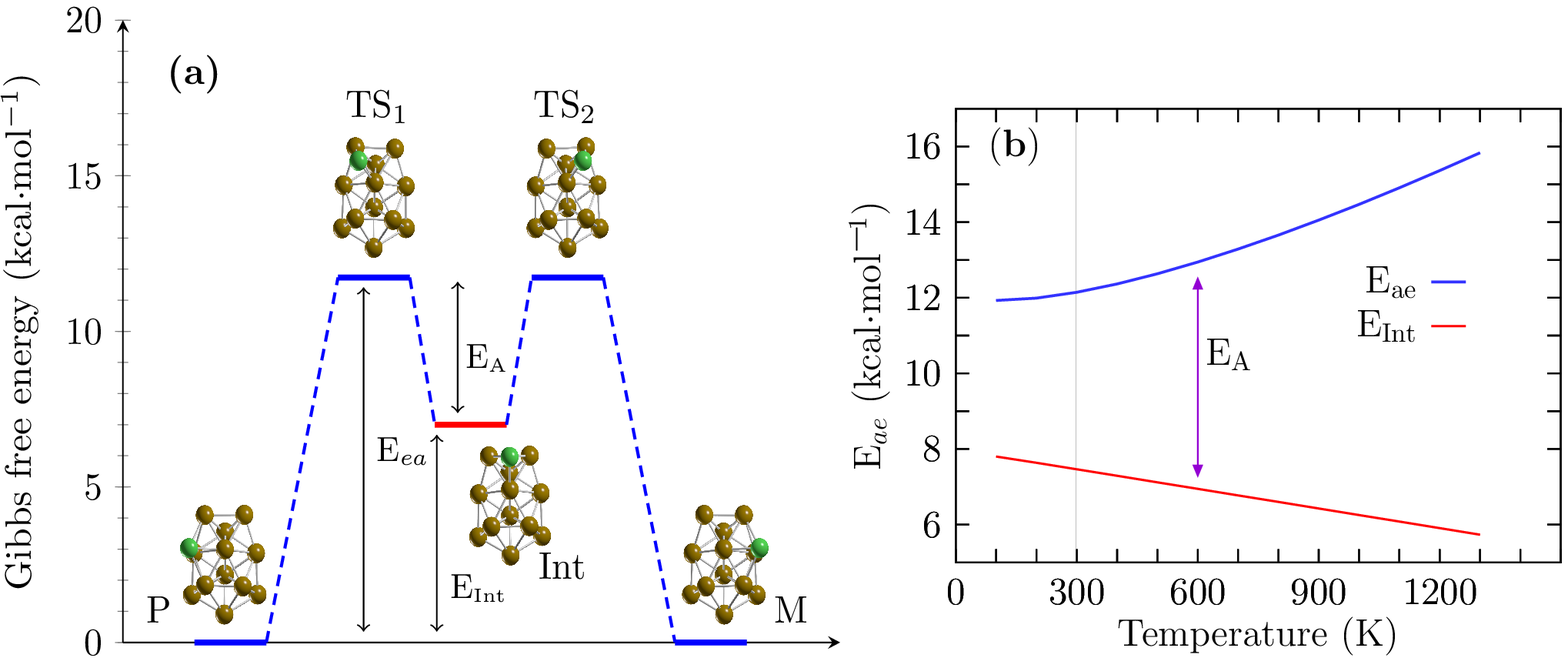}
  \includegraphics[scale=0.75]{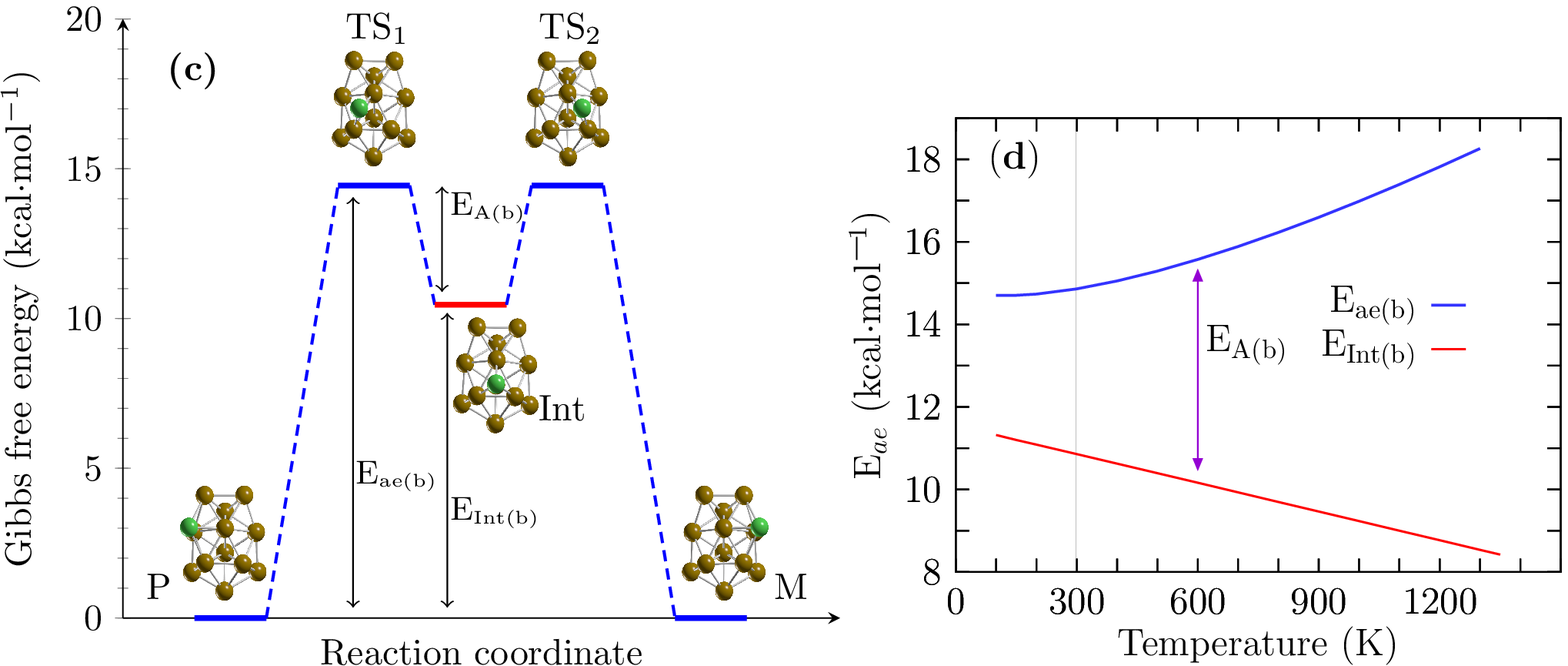}
  \caption{  (Color online)(Color online) Figure (a) shows the energy profile of a chemical reaction (route A) with two symmetric transition states (TS1, TS2) and one intermediate (Int) for the interconversion between the lowest energy $\mathcal{P}$  and $\mathcal{M}$ enantiomers.
 Figure (c) shows the energy profile of a chemical reaction (route B) with two symmetric transition states (TS1, TS2) and one intermediate (Int) for the interconversion between the lowest energy $\mathcal{P}$ and $\mathcal{M}$ enantiomers
 In (b), the enantiomerization energy (Eea) is 12.14 kcal/mol at room temperature. In contrast, the intermediate structure lay at 7.13 kcal/mol (EInt) above the putative global minimum  $\mathcal{P}$ and $\mathcal{M}$  structures, so the activation energy in the interconversion from the intermediate (Int) to the $\mathcal{P}$ or $\mathcal{M}$ structure was 5.0 kcal/mol, at room temperature. Figure (b) displays the enantiomerization energy  (Eea, route A) depicted by a blue-solid line. The relative energy of the intermediate structure (EInt) with respect to the putative global minimum is depicted by a red solid line.   Figure (d) displays the enantiomerization energy  (Eea, route B) depicted by a blue-solid line. The activation energy barrier (EA) between the intermediate structure and the $\mathcal{P}$ or $\mathcal{M}$ structure is indicated by the violet arrow and is the difference between Eea and EInt for temperatures ranging from 100 to 1300 K. The green and copper-colored spheres represent the copper atoms. Green represents a Cu atom that is moving; the chirality is due to this atom.}
\label{barrier}
\end{center}
\end{figure*}
The process in which one enantiomer in a pair is converted into the other is known as enantiomerization;
enantiomers each have an equal probability of occurrence, and the same energy. The enantiomerization energy or
activation energy at temperature T defines the configurational stability. In some cases, a low enantiomerization
energy is undesirable~\cite{https://doi.org/10.1002/chem.202004488}. Two reaction mechanisms compete for the
interconversion from~$\mathcal{P}$ to ~$\mathcal{M}$ structures, and
the shape of the energy barriers (or IRC) is similar to the inverted double-well potential~\cite{Jensen12} 
Figure~\ref{barrier}a shows the reaction mechanism for the interconversion between ~$\mathcal{P}$ and ~$\mathcal{M}$ structures
\href{https://youtu.be/_oqryi3gk3I}{for route A}, 
which proceeds via a two-step mechanism consisting of two symmetric steps with only one intermediate.
Figure~\ref{barrier}a shows the reaction mechanism for the interconversion between ~$\mathcal{P}$ and ~$\mathcal{M}$ structures
\href{https://youtu.be/pMcPvqIMgsw}{for route B}, 
which proceeds via a two-step mechanism consisting of two symmetric steps with only one intermediate.
Figure~\ref{barrier}a depicts the transition states TS1 and TS2, the intermediate (Int), and the putative lowest energy pair of enantiomers
~$\mathcal{P}$ and ~$\mathcal{M}$. 
The energy of enantiomerization was 12.15 kcal/mol, whereas the activation energy for the interconversion of the intermediate to $\mathcal{P}$/$\mathcal{M}$
structures was 5 kcal/mol at room temperature.
The intermediate state was located at 7.13 kcal/mol above the putative chiral lowest energy structure The structures of the TS1 and TS2 states
are depicted in Figure~\ref{barrier}a. They appeared to be bilayer structures composed of a shared pentagonal bipyramid interspersed with a distorted hexagonal ring.
The green atom represents the Cu atom that caps one edge of the pentagonal bipyramid and is responsible for the chirality of the Cu$_{13}$
cluster. The intermediate state structure for the same bilayer presented 12 Cu atoms and the green Cu atom caps
one of the faces in the pentagonal bipyramid. Figure~\ref{barrier}a shows the enantiomerization
energy E$_{ae}$ depicted by a solid blue line. The relative energy of the intermediate, E$_{Int}$, for the putative global chiral
structures is depicted by a solid red line. As the temperature increased, the enantiomerization energy increased almost linearly.
In contrast, the relative energy of the intermediate with respect to the putative global minimum decreased linearly, implying that
the inverted double-wall became energetically greater. The activation energy for the interconversion between the intermediate and the
$\mathcal{M}$ structure was 5 kcal/mol at room temperature; this increased linearly, from 4 kcal/mol at a temperature of 100 K to 9.5 kcal/mol at
a temperature of 1200 K. As a consequence. In contrast, at low temperatures, the enantiomerization energy
trend reached a minimum, whereas the relative energy of the intermediate
increased; thus, the energy activation for the interconversion of the intermediate to the $\mathcal{P}$/$\mathcal{M}$ states tended to be smaller.
These results suggest that at high temperatures, the enantiomerization barrier energy increased, and the intermediate state energy became
more significant, stabilizing chirality and allowing the  separation of enantiomers at room temperature.
To elucidate the behavior of the interconversion from ~$\mathcal{P}$ and ~$\mathcal{M}$ structures;
We computed the reaction rate constants based on Equation~\ref{eyring} (Eyring equation) that used the activation barrier
$\Delta$G between the putative global minimum $\mathcal{P}$/$\mathcal{M}$ structures and the transition state and
did not take into account the tunneling effect. The Eyring equation relates the rate constant to temperature and the activation free energy.
\begin{equation}
\displaystyle
k=k_0\frac{K_BT}{h} e^{-{\Delta G}/{RT}}
\label{eyring}
\end{equation}

\begin{table*}[ht!]
 \centering
 \begin{tabular}{lcccccc}
    \hline \\[-3.8ex]
   \hline \\[-1.8ex]
& \multicolumn{3}{c}{Reaction  A} & \multicolumn{3}{c}{Reaction B}
  \\
   \cline{2-4}\cline{5-7}
Level of theory       & E$_{ea}$  &  E$_{Int}$  & E$_{A}$  & E$_{ea}$  & E$_{Int}$  &  E$_{A}$ \\
\hline \\[-1.8ex]
B3PW91-D3/Def2TZVP   & 12.15    &   7.13     &   5.0    &   14.85   &   10.86   &  3.99\\
B3PW91/Def2TZVP       & 12.36    &   6.38     &   5.97   &   15.60   &   11.31   &  4.28\\
\hline \\[-1.8ex]
\end{tabular}
\caption{Table shows the values of enantiomerization energy ( E$_{\textrm{ea}}$), relative energy of intermediate ( E$_{\textrm{Int}}$), and  activation energy (E$_{\textrm{A}}$), for two mechanism of reaction A and B, and with and witout taking into account the D3 Grimme's dispersion.
  The inclusion of dispersion lower the energy barriers, i.e in B the E$_{\textrm{A}}$ decreases 7{\%}. (from 4.28 to 3.99 kcal/mol)}
\label{tabla2} 
\end{table*}
In Equation~\ref{eyring}, $k$ is the rate constant, $k_0$ transmission coefficient that in the absence of other kinetic data is set to 1,
$K_B$ is Boltzmann constant, T is the temperature, $h$ is the Planck constant, $R$ is ideal gas constant, and $\Delta$G is the activation energy barrier.
We consider the rate-determining step in the overall reaction is the rate of interconversion between  $\mathcal{P}$/$\mathcal{M}$
and intermediary structures, and it is the slowest step; besides, its high activation energy characterizes it.
(Notice, the activation energy barrier $\Delta$G is computed with the statistical thermodynamics).
The height of the activation energy barrier at room temperature for interconversion between $\mathcal{P}$/$\mathcal{M}$ and intermediary
structures in route A is 12.14 kcal/mol, which leads to a rate constant of 7.84$\times$103 1/s, whereas the activation energy barrier at 900 K is
14.0 kcal/mol, which leads to a rate constant of 7.47$\times$109 1/s. This show that the rate constant increases at high temperatures,
and it agrees with the thermal populations where the contribution of all isomers is less than 10{\%} at high temperatures.
We also have to consider that the melting point for copper is 1358 K, thus around this temperature, the glass state will dominate.
Regarding dispersion, if it is not considered, the energy barriers tend to increase. For ease of comparison, Table 2 shows the values of
the two similar reaction mechanisms A and B, taking into account the D3 dispersion of Grimme.
Energetically, the reaction mechanism of route B is not so different from that of route A, as we can see in Figure~\ref{barrier}c,d and in Table~\ref{tabla2}.

\section{Conclusions}
For the first time, to our knowledge and from our results, we computed the effect of symmetry on the Boltzmann populations on TM Cu clusters.
Our computed $\mathcal{T}_1$ diagnostic 0.023,  determine that the computed DFT energies are not properly described by a single reference method  or contain a multireference character. So, further studies needed to be done.
We explored the potential and free energy surface of the neutral Cu$_{13}$ cluster with an efficient cascade-type algorithm coupled to DFT. We found that the putative global minimum was a pair of enantiomers that strongly dominated at room temperature. Our findings show that the chirality exhibited by the Cu$_{13}$
cluster emerged from the Cu atom capping a face of the core Cu$_{12}$ cluster. We showed that for the interconversion between $\mathcal{P}$ and $\mathcal{M}$
structures, two similar reaction mechanisms were possible. Both of them closed in their energy barriers and proceeded via two symmetric steps. The energy of enantiomerization and the energy barrier between the intermediate and the $\mathcal{P}$/$\mathcal{M}$ structures increased as the temperature increased. We computed the reaction rate constants based on the Eyring equation; our findings show that at high temperatures, enantiomerization is favored. The entropic-temperature term did not significantly influence the energy barriers; thus, they are mainly composed of enthalpic energy. Regarding Grimme’s dispersion D3, this lowers the energy barriers, i.g., in route B, the EA decreased by 7{\%} (from 4.3 to 4.0 kcal/mol). We showed that the pair of enantiomers with C$_1$ symmetries strongly dominated at room temperature revealed by the thermal population. Hence, at body temperature, all the molecular properties were attributable to those structures. For each DFT functional (B3PW91, TPSS, PBE, and BP86) used in this study, the thermal population and energetic ordering of the isomers are preserved, although differences in the energy between the isomers occur; the main contributors to any molecular property of Cu$_{13}$ are always the chiral isomers. The bonding in the lowest energy chiral Cu$_{13}$ cluster is due to the 4s shell electrons for which the bonding pattern, as revealed by AdNDP, consisted of 6 sets of 13c-2e completely delocalized bonds, plus a 9c-1e bond corresponding to the unpaired electron. Future work will focus on the computation of UV absorption of the Cu clusters employing Boltzmann weighted spectra, comparing it with a single UV spectrum of the putative global minimum.

\section{Acknowledgments}
C. E. B.-G. thanks Conacyt for the Ph.D. scholarship (860052). E. P.-S. thanks Conacyt for the Ph.D. scholarship (1008864).  E. R.-CH. thanks Conacyt for the Ph.D. scholarship (1075701). We are grateful to Dra. Carmen Heras, and L.C.C. Daniel Mendoza for granting us access to their clusters and computational support. Computational resources for this work were provided by \emph{ACARUS} through the High-Performance Computing Area of the University of Sonora, Sonora, M\'exico. We are also grateful to the computational chemistry laboratory for providing computational resources, \emph{ELBAKYAN}, and \emph{PAKAL} supercomputers. Powered{@}NLHPC: This research was partially supported by the supercomputing infrastructure of the NLHPC (ECM-02) in Chile.
\section{Conflicts of Interest} The authors declare no conflict of interest.
\section{Funding} This research received no external funding.
\section{Supplementary Materials}
The following are available online \href{https://youtu.be/_oqryi3gk3I}{for route A}, and \href{https://youtu.be/pMcPvqIMgsw}{for route B}.
All xyz atomic coordinates optimized of the Cu$_{13}$B$_{8}$ cluster at the B3PW91-D3/def2-TZVP/Freq.
\section{Abbreviations}
The following abbreviations are used in this manuscript:\\
Density Functional Theory (DFT), Domain-based Local Pair Natural\ Orbital Coupled-Cluster Theory (DLPNO-CCSD(T)),
Zero-Point Energy (ZPE), Global Genetic Algorithm of University of Sonora (GALGOSON), Adaptive Natural Density Partitioning (AdNDP).  
\bibliographystyle{unsrt}
\bibliography{bibliografia}
\newpage 
\appendix
\newpage
\appendix
\onecolumngrid
\section{Boltzmann Probabilities computed at TPSS/def2-TZVP PBE/def2-TZVP PB68/def2-TZVP}
\label{appendix:a}
\begin{figure}[ht!]
\begin{center}  
\includegraphics[scale=0.85]{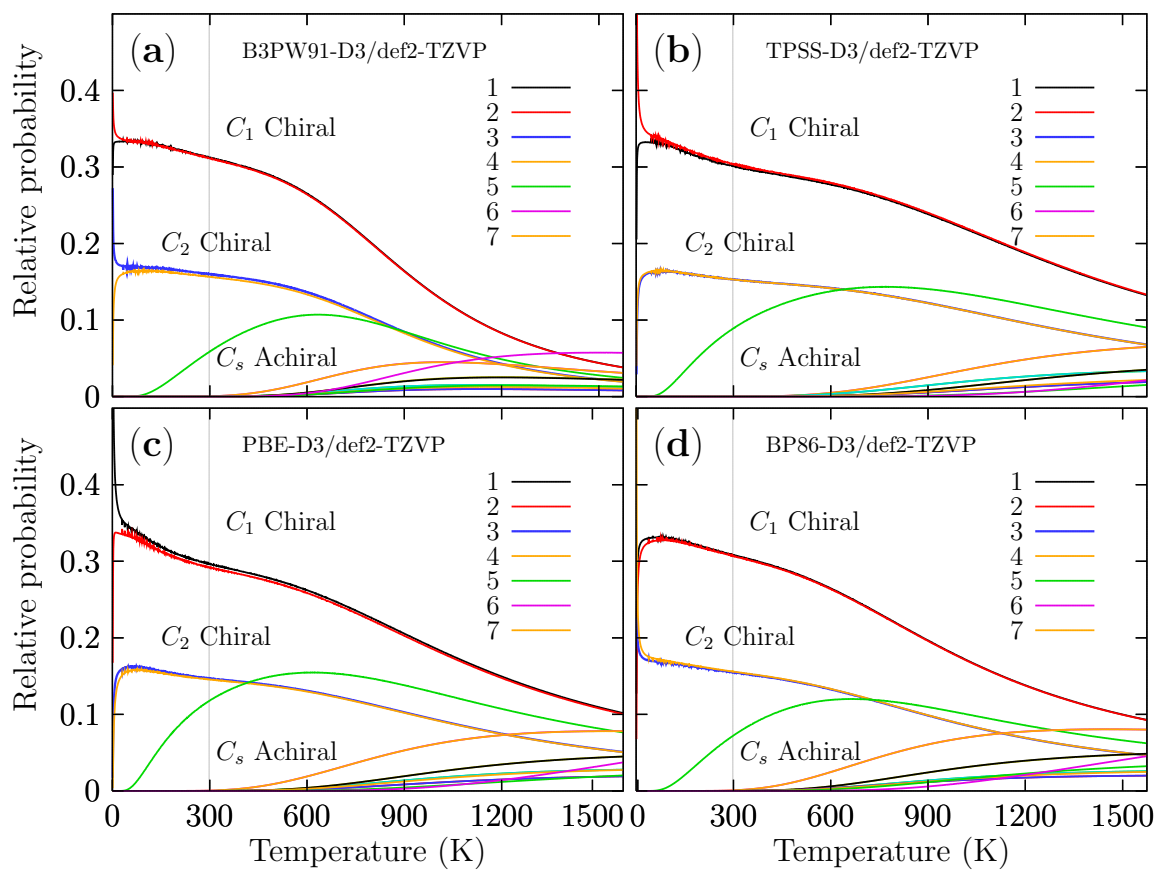}
\caption{(Color online) Probability of occurrence for all isomers at temperatures ranging from 20 to 1500 K computed with  DFT  functionals: TPSS, PBE, and BP86 The bulk melting temperature of copper is 1358 K~\cite{Grigoryan}; thus, our results below this temperature are consistent.} 
\label{popu2}
\end{center}
\end{figure}
\newpage
\section{XYZ atomic coordiantes}
\label{appendix:b}
\begin{verbatim}
13    
0.0         
Cu  -0.990009000000   0.202990000000   0.910273000000
Cu   0.989148000000   0.203204000000  -0.910494000000
Cu   1.307754000000   0.817211000000   1.506707000000
Cu  -0.756524000000  -1.521352000000  -1.012184000000
Cu  -2.995439000000  -0.494575000000  -0.267184000000
Cu  -0.000551000000   2.308713000000  -0.000120000000
Cu   1.705678000000  -2.128579000000  -1.124758000000
Cu   2.995766000000  -0.493776000000   0.267529000000
Cu   0.756822000000  -1.521006000000   1.012214000000
Cu  -1.308834000000   0.816218000000  -1.506262000000
Cu  -1.705107000000  -2.129263000000   1.124856000000
Cu   2.516484000000   1.969366000000  -0.285914000000
Cu  -2.515188000000   1.970848000000   0.285338000000
13 
0.0                
Cu   0.000000000000   1.344973000000   0.202955000000
Cu   0.000000000000  -1.344973000000   0.202955000000
Cu  -1.995454000000   0.058043000000   0.817297000000
Cu   1.256550000000  -0.129438000000  -1.520715000000
Cu   2.225189000000   2.023127000000  -0.494441000000
Cu   0.000000000000   0.000000000000   2.309851000000
Cu  -0.329078000000  -2.016616000000  -2.128802000000
Cu  -2.225189000000  -2.023127000000  -0.494441000000
Cu  -1.256550000000   0.129438000000  -1.520715000000
Cu   1.995454000000  -0.058043000000   0.817297000000
Cu   0.329078000000   2.016616000000  -2.128802000000
Cu  -1.493610000000  -2.046117000000   1.968780000000
Cu   1.493610000000   2.046117000000   1.968780000000
\end{verbatim}
\typeout{get arXiv to do 4 passes: Label(s) may have changed. Rerun}
\end{document}